\documentclass[a4paper]{article}

\usepackage{INTERSPEECH2022}

\usepackage{graphicx}
\usepackage{amssymb,amsmath,bm}
\usepackage{textcomp}
\usepackage{url}
\usepackage[utf8]{inputenc}
\usepackage{CJKutf8}
\usepackage{multirow}
\usepackage{subfigure}
\usepackage{comment}
\usepackage{multirow}

\newcommand{\Fig}[1]{Fig. \ref{fig:#1}} 

\renewcommand{\Vec}[1]{\textrm{\boldmath $#1$}} 

\newcommand{\x}{ \Vec{x} } 
\newcommand{\y}{ \Vec{y} } 
\newcommand{\p}{ \Vec{p} } 
\newcommand{\lng}{ \Vec{l} } 

\ninept

\title{Acoustic Modeling for End-to-End Empathetic Dialogue Speech Synthesis Using Linguistic and Prosodic Contexts of Dialogue History}

\makeatletter
\def\name#1{\gdef\@name{#1\\}}
\makeatother
\name{{\em Yuto Nishimura$^{1}$, Yuki Saito$^{1}$, Shinnosuke Takamichi$^1$, Kentaro Tachibana$^2$, and Hiroshi Saruwatari$^1$}}

\address{
$^1$The University of Tokyo, Japan, $^2$LINE Corp., Japan.
}
\email{
yutonishimurav20512@gmail.com, yuuki\_saito@ipc.i.u-tokyo.ac.jp
}

\begin{document}
\setlength{\abovedisplayskip}{5pt} 
\setlength{\belowdisplayskip}{5pt} 
\setlength\floatsep{3pt} 
\setlength\intextsep{5pt} 
\setlength\textfloatsep{10pt} 
\setlength{\dbltextfloatsep}{6pt} 
\setlength{\dblfloatsep}{2pt}

\maketitle

\begin{abstract}
We propose an end-to-end empathetic dialogue speech synthesis (DSS) model that considers both the linguistic and prosodic contexts of dialogue history. Empathy is the active attempt by humans to get inside the interlocutor in dialogue, and empathetic DSS is a technology to implement this act in spoken dialogue systems. Our model is conditioned by the history of linguistic and prosody features for predicting appropriate dialogue context. As such, it can be regarded as an extension of the conventional linguistic-feature-based dialogue history modeling. To train the empathetic DSS model effectively, we investigate 1) a self-supervised learning model pretrained with large speech corpora, 2) a style-guided training using a prosody embedding of the current utterance to be predicted by the dialogue context embedding, 3) a cross-modal attention to combine text and speech modalities, and 4) a sentence-wise embedding to achieve fine-grained prosody modeling rather than utterance-wise modeling. The evaluation results demonstrate that 1) simply considering prosodic contexts of the dialogue history does not improve the quality of speech in empathetic DSS and 2) introducing style-guided training and sentence-wise embedding modeling achieves higher speech quality than that by the conventional method.
\end{abstract}

\section{Introduction}

The role of communication is to understand each other and share thoughts~\cite{komiya15keichou}. \textit{Empathy}, which is the focus of this paper, is a key component of attentive-listening communication. Empathy is the act of feeling  another person's thoughts as if they were the listener's own~\cite{komiya15keichou}, and can be regarded as an active attempt to get inside the other person's head~\cite{davis18empathy}. Unlike \textit{sympathy}, where one synchronizes self with the other in emotion, empathy requires making a response by deeply understanding the other's emotion. Recent studies in dialogue have proposed natural language understanding~\cite{li20empatheticdialogue} and language response generation~\cite{rashkin19empatheticconversation} methods to implement the empathetic behavior in chat-oriented dialogue~\cite{chen17survey_dialogue}. At the same time, emotion and prosody of the response speech are also vital in empathy~\cite{regenbogen12differentialempathy}. The promising method to involve these factors in synthetic speech is called \textit{empathetic dialogue speech synthesis (DSS)}~\cite{saito22studies}. To achieve better empathetic DSS technologies, we need to develop a method that considers the dialogue history with the user (i.e., interlocutor of the dialogue) and generates suitable speech features for the response.

Humans can produce empathetic speech by understanding \textit{contexts} from the linguistic and prosodic features in the dialogue. One possible way to achieve context-aware empathetic DSS technologies is to condition a speech synthesis model by the dialogue history. Guo et al.~\cite{haohan20} proposed a DSS method incorporating a pre-trained language model to obtain \textit{linguistic} features of the dialogue history. Their speech synthesis model is conditioned by phonemes (as with the basic speech synthesis) and a text-chat history's embedding vector extracted from the linguistic features by using a conversational context encoder (CCE). This method can train a DSS model that synthesizes natural speech considering the dialogue context in an end-to-end manner without explicitly using any annotations about the context, such as emotion labels or dialogue situation description. However, it cannot consider the \textit{prosodic} features appearing in the speech history, i.e., the variation of speaking styles in empathetic speech, which is problematic because these can be a crucial means for empathetic communication~\cite{regenbogen12differentialempathy}. Although several studies have attempted to produce natural response speech by using the interlocutor's previous spoken utterance (e.g., Yamazaki et al.'s $F_0$ control method for DSS~\cite{yamazaki21_interspeech}), they are unable to learn the long-term dependencies of spoken dialogue necessary for empathetic prosodic control or to benefit from the powerful recognition-synthesis capabilities of deep learning models.

\begin{figure}[t]
  \centering
  \includegraphics[width=0.8\linewidth, clip]{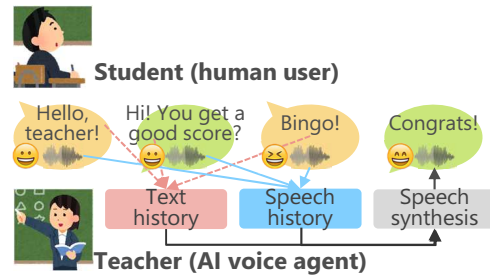}
  \vspace{-5pt}
  \caption{Concept of our work. Speech waveform of current dialogue turn is synthesized considering text and speech history.}
  \label{fig:concept}
\end{figure}

We propose an end-to-end empathetic DSS model that considers both the linguistic and prosody features, as shown in \Fig{concept}. Our model consists of four sub-models: 1) a BERT-based sentence encoder~\cite{devlin19} that extracts text embeddings from a text sequence of dialogue lines, 2) a prosody encoder that extracts prosody embeddings from a mel-spectrogram sequence of the spoken dialogue, 3) a cross-modal (CM) CCE that aggregates the two embeddings and predicts the conversational context from the dialogue history, and 4) a speech synthesis model based on FastSpeech 2~\cite{ren21} conditioned by phoneme embeddings and the context. Therefore, our model can be regarded as an extension of conventional ones~\cite{haohan20,yamazaki21_interspeech} that use only linguistic context information or limited prosody representation in DSS. We also investigate four methods to effectively train our model, which introduce 1) a self-supervised learning (SSL) model pretrained with large speech corpora, 2) a style-guided training using a prosody embedding of a current utterance to be predicted by the dialogue context embedding, 3) a cross-modal attention to capture the text-speech modality, and 4) a sentence-wise embedding to achieve fine-grained prosody modeling rather than utterance-wise modeling. We conducted experiments to determine our model's effectiveness using a corpus of empathetic dialogue speech synthesis~\cite{saito22studies}. The results demonstrate that 1) simply considering prosodic contexts of the dialogue history does not improve the quality of speech by empathetic DSS and 2) introducing style-guided training and sentence-wise embedding modeling achieves higher speech quality than that by the conventional method.

\vspace{-5pt}
\section{Related Work}
\vspace{-3pt}
\subsection{End-to-end speech synthesis}
\vspace{-3pt}
End-to-end speech synthesis~\cite{wang17tacotron} is a framework to solve three sub-tasks in speech synthesis (i.e., text analysis, acoustic modeling, and waveform synthesis) using a stack of deep neural networks (DNNs). The definition of this framework depends on where we consider the end to be, e.g., linguistic features to waveform~\cite{oord16wavenet}, text (phonemes) to acoustic features~\cite{wang17tacotron,shen18}, or text to waveform (i.e., truly end-to-end methods)~\cite{sotelo17,ping19,weiss21}. In this study, we follow the second definition and adopt FastSpeech 2~\cite{ren21} as a baseline model from the viewpoints of fast learning speed and stable inference.

\vspace{-3pt}
\subsection{Dialogue-context-aware acoustic modeling}
\vspace{-3pt}
Context-aware acoustic modeling for end-to-end DSS has to deal with the difficulty of modeling the conversational context with long-term dependency of the dialogue (i.e., dialogue history). Guo et al.~\cite{haohan20} proposed a DSS method introducing a CCE that predicts the contexts from a text sequence of dialogue history, without using any annotations related to the dialogue, such as emotion labels~\cite{chiba18sigdial}. However, their method cannot consider prosodic features of the dialogue history for the context prediction, which is unfortunate because prosody is often vital for empathetic communication. Yamazaki et al.~\cite{yamazaki21_interspeech} proposed a spoken-response generation method that can control the $F_0$ values of the response speech while considering both the linguistic and prosodic contexts of the user's previous utterance. However, this method focuses only on the pitch control and ignores other prosodic contexts such as energy, which is also an essential factor to simulate the empathetic behavior in human speech communication~\cite{saito22studies}. Cong et al.~\cite{cong21} proposed a style-guided context learning for a DSS model, where embedding vectors of the current utterance by a dialogue agent and previous utterance by a user extracted from their mel-spectrograms become closer in the embedding space. However, this method cannot consider the complicated interaction between the linguistic and prosodic contexts of the dialogue history in empathetic spoken dialogue.

\vspace{-3pt}
\subsection{Empathetic dialogue generation}
\vspace{-3pt}
In daily communication, humans naturally control their speech prosody to empathize with each other. Empathetic dialogue generation aims to incorporate this ``empathetic dialogue'' behavior into a dialogue system by implementing empathy in human communication, which can improve the user experience in human-computer interactions~\cite{liu05}. A major challenge in building such systems is how to estimate the user's mental state (typically expressed as emotions), from the dialog history. Recently developed DNN-based natural language understanding and generation models have the ability to deeply understand the user's mental state, intentions, and emotions~\cite{liu21iwsds,xie21conll}. However, context-aware empathetic DSS technology has not been achieved thus far.

\begin{figure}[t]
  \centering
  \includegraphics[width=0.95\linewidth, clip]{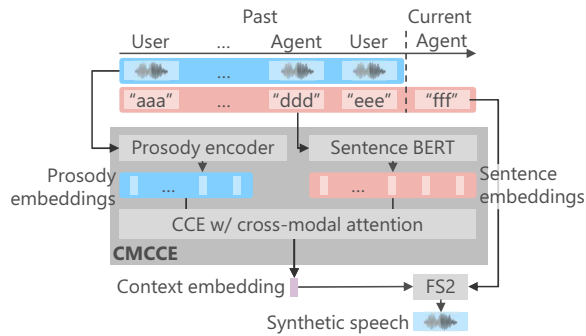}
  \vspace{-5pt}
  \caption{Overview of proposed method. ``FS2'' stands for ``FastSpeech2.''}
  \label{fig:proposed}
\end{figure}


\vspace{-5pt}
\section{Proposed Method}
\vspace{-3pt}
In this study, we propose a speech synthesis method conditioned on the linguistic and prosodic contexts of dialogue history between the agent and user. \Fig{proposed} shows an overview of the proposed method.

\vspace{-3pt}
\subsection{Baseline empathetic DSS model}\label{subsect:baseline}
\vspace{-3pt}
Our baseline empathetic DSS model consists of a BERT-based sentence encoder~\cite{devlin19}, CCE~\cite{haohan20}, and speech synthesis model based on FastSpeech 2~\cite{ren21}. Let $C$ be a memory capacity parameter that determines the number of past utterances by the agent and user to be considered for the context prediction. In generating a dialogue response at the $t$th turn, the sentence BERT first takes a text sequence of the dialogue up to the past $C$ turns, i.e., $\x_{t-C}, \x_{t-C+1}, \ldots, \x_{t}$, and produces their text embeddings $\lng_{t-C}, \lng_{t-C+1}, \ldots, \lng_{t}$. Note that vectors with subscript $t-C < 1$ are treated as zero vectors and never contribute to the context prediction. Henceforth, we denote these sequences as $\x_{t-C:t}$ and $\lng_{t-C:t}$. The CCE then extracts a conversational context embedding vector $\Vec{e}_{t}$ from $\lng_{t-C:t}$. Finally, the FastSpeech 2 model takes the text to be synthesized at turn $t$, $\x_{t}$, and the embedding vector, $\Vec{e}_{t}$, as input and generates a mel-spectrogram of response speech ${\hat \y}_{t}$. The loss function for training this DSS model is defined as the speech parameter prediction error (e.g., L1 loss) between ${\hat \y}_{t}$ and $\y_{t}$, i.e., the mel-spectrogram extracted from the agent's speech at the $t$th turn.

\vspace{-3pt}
\subsection{Cross-modal CCE (CMCCE)}\label{subsect:cmcce}
\vspace{-3pt}
We introduce a CMCCE, an extension of the conventional CCE, to our empathetic DSS model so that it can consider both the linguistic and prosodic contexts of long-term dialogue history. Figure~\ref{fig:proposed} shows the architecture of our CMCCE. In the conversational context prediction using CMCCE, a DNN-based prosody encoder first takes a mel-spectrogram sequence of the dialogue up to the past $C$ turns excluding the current turn $t$, i.e., $\y_{t-C:t-1}$, and produces their prosody embeddings $\p_{t-C:t-1}$. Then, a sentence BERT extracts the text embeddings $\lng_{t-C:t}$ from a dialogue text sequence $\x_{t-C:t}$ in the same manner as described in Section~\ref{subsect:baseline}. Finally, our CMCCE aggregates the two embeddings $\p_{t-C:t-1}$ and $\lng_{t-C:t}$ to predict the context embedding $\Vec{e}_{t}$.

In the following subsections, we present four strategies to make our CMCCE capture the linguistic and prosodic contexts of dialogue history effectively.
\vspace{-3pt}
\subsubsection{SSL model as prosody encoder}
\vspace{-3pt}
One can train a random-initialized DNN (e.g., Du et al.'s work~\cite{du22taslp}) from scratch. Another approach is to use an SSL model pretrained with large speech corpora as the prosody encoder and fix it during the training, similar to the sentence BERT used for the text-related embedding extraction.

\vspace{-3pt}
\subsubsection{Style-guided context embedding learning}
\vspace{-3pt}
The context embedding $\Vec{e}_{t}$ should correspond to the appropriate speaking style for the text at the $t$th turn to be synthesized. To explicitly consider this requirement, we introduce the style-guided context embedding learning proposed by Cong et al.~\cite{cong21} into empathetic DSS. Specifically, we train our empathetic DSS model to minimize not only the speech parameter prediction error (described in Section~\ref{subsect:baseline}) but also the embedding matching loss defined as the mean squared error between the context embedding $\Vec{e}_{t}$ and prosody embedding $\Vec{p}_{t}$. Specifically, we train the DSS model so that the embedding obtained from the past dialogue history matches the embedding obtained from the ground-truth speech at the $t$th turn. The difference between our method and Cong et al.'s is the dialogue history information to be used for the context prediction. In other words, the former uses the linguistic and prosodic context with long-term history while the latter only considers the prosody embedding of a user's utterance as the context.

\begin{figure}[t]
  \centering
  \includegraphics[width=0.8\linewidth, clip]{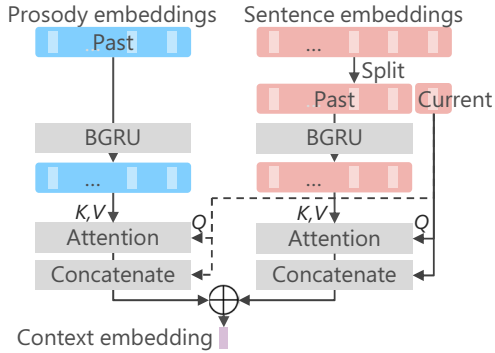}
  \vspace{-5pt}
  \caption{Attention-based cross-modal aggregation architecture. Blue and red boxes represent prosody and text embeddings, respectively. $Q, K$, and $V$ are query, key, and value of attention, respectively.}
  \label{fig:attn}
\end{figure}
\vspace{-3pt}
\subsubsection{Attention-based cross-modal aggregation}
\vspace{-3pt}
One simple approach to aggregate the embeddings from two different modalities, i.e., text and speech, is to compress each embedding sequence into a fixed-dimensional embedding independently, and take the element-wise sum of the two outcomes. However, this approach does not necessarily model the cross-modality of text and speech in empathetic dialogue history. Therefore, we investigate another approach to aggregate the cross-modal embeddings by using attention networks, as shown in \Fig{attn}. In this approach, the text embedding sequence $\lng_{t-C:t}$ is first split into the current and past embeddings, $\lng_{t}$ and $\lng_{t-C:t-1}$. Then, the past text/prosody embedding sequences $\lng_{t-C:t-1}$ and $\p_{t-C:t-1}$ are fed into a bi-directional gated recurrent unit (BGRU) independently to capture their time dependency. The outcomes of this operation are regarded as the key and value of the following attention networks, and the current text embedding $\lng_{t}$ is treated as a query of the attention. These attention networks can be expected to learn which point in the past information should be used for predicting the conversational context while considering both text and speech modalities.

\vspace{-3pt}
\subsubsection{Fine-grained context embedding modeling}
\vspace{-3pt}
Guo et al.'s conventional CCE predicts an {\it utterance-wise} context embedding, i.e., it assumes that speech utterance in each turn has only one speaking style. However, empathetic speech is a bit more complicated, as an agent typically considers a user's mental state and then responds to the user with one or more speaking styles in the same turn. Therefore, we instead use {\it sentence-wise} context embeddings for modeling fine-grained change of the speaking style in empathetic dialogue within a single utterance. Note that this fine-grained modeling is only applied to generation for the current sentence, and utterance-wise embeddings of past information about the dialogue are stored in the dialogue history. The reason is to prevent the dialogue history from being filled only with the sentence-wise embeddings of one speaker (i.e., agent or user).

\begin{table}[t]
\centering
\caption{Number of dialogues in training/validation/evaluation sets for experimental evaluation. Numbers in parentheses mean the number of utterances by the teacher.}
\label{tab:teacher_data_split}
\vspace{-3mm}
\begin{tabular}{c|rrrr}
\hline
\hline
Set & & Long & & Short \\
\hline
Training & 126 & (1,743) & 600 & (1,200) \\
Validation & 12 & (91) & 60 & (120) \\
Evaluation & 12 & (91) & 60 & (120) \\
\hline
\hline
\end{tabular}
\end{table}

\vspace{-5pt}
\section{Experimental Evaluation}
\vspace{-3pt}
\subsection{Experimental conditions}
We used the STUDIES corpus~\cite{saito22studies}, which contains simulated empathetic chat-dialogue uttered by three actors playing the roles of a female teacher and her male/female students at a tutoring school, for the experimental evaluation. We built the empathetic DSS model of the teacher in the STUDIES corpus. We split the teacher's speech data included in the long dialogue (10--20 turns) and short dialogue (four turns) subsets for training, validation, and evaluation, as shown in Table~\ref{tab:teacher_data_split}. We downsampled the speech data to 22,050~Hz.

We used FastSpeech 2~\cite{ren21} with the PyTorch implementation for Japanese speech synthesis~\cite{fastspeech2jp}. We followed the default settings of the DNN architecture and speech parameter extraction of this implementation. We used the WORLD vocoder~\cite{morise16world,morise16d4c} to estimate $F_0$. We modified the variance adaptors to predict the speech features averaged over phonemes, referring to FastPitch~\cite{lancucki21} for stable training and higher synthetic speech quality. The optimization algorithm was Adam~\cite{kingma14adam} with an initial learning rate $\eta$ of 0.0625, $\beta_1$ of 0.9, and $\beta_2$ of 0.98. We first pretrained FastSpeech 2 using the JSUT corpus~\cite{takamichi20ast} for 500 epochs and then fine-tuned it using the STUDIES corpus for 500 epochs.

We used the HiFi-GAN vocoder~\cite{kong20} for speech waveform generation from 80-dimensional mel-spectrograms. We first pretrained HiFi-GAN using the JSUT corpus~\cite{takamichi20ast} for 100 epochs. Then, we fine-tuned it by using the same training data as that for FastSpeech 2 (shown in Table~\ref{tab:teacher_data_split}) for 200 epochs. The optimization algorithm was Adam with an initial learning rate $\eta$ of 0.0003, $\beta_1$ of 0.8, and $\beta_2$ of 0.99.

We used the CCE~\cite{haohan20} with the open-source PyTorch implementation~\cite{cce_github}. However, since the open-sourced implementation of context
aggregation was different from that in the original paper, we changed the structure to be the same as in the original paper. In addition, although Guo et al.'s method also used an auxiliary encoder to consider hand-crafted linguistic features, we excluded it in this study so as to focus only on the context embedding prediction from the dialogue history information. We used a sentence BERT pretrained using Japanese text data\footnote{https://huggingface.co/colorfulscoop/sbert-base-ja} for the text embedding extraction in both the baseline and proposed models. For the pretrained prosody encoder, we used the wav2vec 2.0 SSL model\footnote{https://huggingface.co/jonatasgrosman/wav2vec2-large-xlsr-53-japanese}~\cite{baevski20} pretrained with Japanese speech data in Common Voice~\cite{ardila20}, CSS10~\cite{park19css10}, and JSUT~\cite{takamichi20ast} corpora. In our DSS method, the prosody embeddings were extracted from the agent's ground-truth speech and synthetic speech for the training and inference phases, respectively.

\vspace{-3pt}
\subsection{Subjective evaluations}
\vspace{-3pt}

We conducted preference AB tests on the naturalness of synthetic speech by the ``\textbf{Baseline}'' model (FastSpeech 2 conditioned by text-embedding-derived context embedding~\cite{haohan20}) and our empathetic DSS model with the four additional strategies (described in Section~\ref{subsect:cmcce}). Similarly, we conducted preference XAB tests on the speaking-style similarity of synthetic speech to the teacher's corresponding speech sample analysis-synthesized by the HiFi-GAN vocoder. In the XAB test, we first asked participants to listen to sample ``X'' (analysis-synthesized speech) and then to choose which speaking style (way of expressing emotion) was closer to ``X'' among the ``A'' and ``B'' samples  that were played afterward. Twenty-five listeners participated in each of the evaluations using our crowdsourcing evaluation system. Each listener evaluated ten samples. In this evaluation, we first verified the effectiveness of the four strategies (Section~\ref{subsect:cmcce}) to be considered individually. Then, for those that showed positive effect on the quality of synthetic speech, we further evaluated the interaction of multiple strategies.

Table~\ref{tab:ab_xab} shows the evaluation results, where ``\textbf{SG},'' ``\textbf{Attn},'' and ``\textbf{FG}'' stand for style-guided embedding learning, attention for cross-modal aggregation, and fine-grained context modeling for our empathetic DSS model, respectively. For the first investigation results, we found that just considering the speech modality in empathetic DSS did not necessarily improve the quality of synthetic speech. At the same time, introducing one of our strategies, i.e., ``SG'' for the proposed model without the SSL model and ``Attn'' for that with the SSL model, could outperform the baseline model regarding the speech quality. Furthermore, the proposed model with multiple strategies, i.e., without the SSL model and with the ``SG'' and ``FG'' strategies, achieved significantly higher scores than the baseline in both naturalness and speaking-style similarity. These results suggest that it may be possible to estimate the dialogue context that can achieve higher quality empathetic DSS by considering not only textual features but also prosodic ones of the dialogue history. It should be noted, however, that inappropriate strategies (e.g., ``FG'' for the proposed model without the SSL model) can degrade the quality of the synthesized speech over the baseline. One reason may be the difficulty of learning complicated contexts with insufficient modeling performance.

Finally, we conducted a five-scale mean opinion score (MOS) test on the utterance-level naturalness of synthetic speech to evaluate the absolute performance of the baseline and proposed models. In this test, we compared the baseline model with the best proposed model (i.e., without SSL model and with ``SG'' and ``FG'' strategies) and its variants for an ablation study. We presented 35 synthetic speech samples to listeners in random order. Listeners rated the naturalness of each sample from degrees of 1 (``very unnatural'') to 5 (``very natural'').

Table~\ref{tab:mos} shows the evaluation results. Among the seven compared methods, our model introducing ``SG'' and ``FG'' without the SSL model achieved the highest MOS, thus demonstrating the effectiveness of our model in improving the quality of synthetic speech. Interestingly, the proposed method that introduced all strategies excluding ``SSL model'' performed similarly to the baseline. This result suggests that in empathetic DSS, using a richer model does not necessarily lead to an improvement in speech quality. The speech samples used in the subjective evaluations are available at http://sython.org/Corpus/STUDIES/demo\_empDSS.html.

\begin{table}[tb]
    \centering
    \footnotesize
    \caption{Results of preference AB and XAB tests. {\bf BOLD} values have significant difference with $p < 0.05$ (Baseline vs Proposed method with some of techniques described in Section~\ref{subsect:cmcce})}
    \label{tab:ab_xab}
    \vspace{-3mm}
    \subtable[Results of proposed model without SSL model]{
    \begin{tabular}{c|c|c|ccc}
    \hline\hline
    & \multirow{2}{*}{Naturalness} & \multirow{2}{*}{Similarity} & \multicolumn{3}{|c}{Proposed (w/o SSL)} \\
    Baseline & & & SG & Attn & FG \\
    \hline
    & \textbf{0.45 vs. 0.55} & \textbf{0.54 vs. 0.46} & & & \\
    & \textbf{0.44 vs. 0.56} & 0.53 vs. 0.47 & \checkmark  & & \\
    & 0.50 vs. 0.50 & 0.54 vs. 0.46 & & \checkmark & \\
    & 0.48 vs. 0.52 & \textbf{0.54 vs. 0.46 }& & & \checkmark \\
    \hline
    & \textbf{0.43 vs. 0.57} & 0.47 vs. 0.53 & \checkmark & \checkmark & \\
    & \textbf{0.45 vs. 0.55} & \textbf{0.45 vs. 0.55} & \checkmark & & \checkmark \\
    \hline\hline
    \end{tabular}
    }
    \subtable[Results of proposed model with SSL model]{
    \begin{tabular}{c|c|c|ccc}
    \hline\hline
    & \multirow{2}{*}{Naturalness} & \multirow{2}{*}{Similarity} & \multicolumn{3}{|c}{Proposed (w/ SSL)} \\
    Baseline & & & SG & Attn & FG \\
    \hline
    & 0.50 vs. 0.50 & \textbf{0.61 vs. 0.39} & & & \\
    & 0.53 vs. 0.47 & 0.46 vs. 0.54 & \checkmark  & & \\
    & 0.51 vs. 0.49 & \textbf{0.44 vs. 0.56} & & \checkmark & \\
    & 0.52 vs. 0.48 & 0.50 vs. 0.50 & & & \checkmark \\
    \hline
    & \textbf{0.43 vs. 0.57} & 0.50 vs. 0.50 & \checkmark & \checkmark & \\
    & 0.46 vs. 0.54 & \textbf{0.54 vs. 0.46} & & \checkmark & \checkmark \\
    \hline\hline
    \end{tabular}
    }
\end{table}

\begin{table}[tb]
    \centering
    \footnotesize
    \caption{Results of MOS tests on speech naturalness with 95\% confidence intervals}
    \label{tab:mos}
    \vspace{-3mm}
    \begin{tabular}{cccc}
    \hline\hline
    \multicolumn{3}{c}{Method} & Naturalness MOS \\
    \hline
    \multicolumn{3}{c}{Proposed (w/o SSL)} & \\
    SG & Attn & FG & \\
    \hline
    & & & 3.59$\pm$0.10 \\
    \checkmark & & & 3.62$\pm$0.10 \\
    & & \checkmark & 3.59$\pm$0.10 \\
    \checkmark & & \checkmark & 3.66$\pm$0.10 \\
    \checkmark & \checkmark & \checkmark & 3.55$\pm$0.10 \\
    \hline
    \multicolumn{3}{c}{Baseline} & 3.55$\pm$0.10 \\
    \hline\hline
    \end{tabular}
\end{table}

\vspace{-5pt}
\section{Conclusion}
\vspace{-3pt}
In this paper, we proposed an end-to-end empathetic DSS model that considers both linguistic and prosody features, and investigated the effectiveness of four methods to train our model. The results demonstrated that 1) simply considering  the prosodic contexts of the dialogue history did not improve the quality of speech by empathetic DSS and 2) introducing style-guided training and sentence-wise embedding modeling achieved higher speech quality than that by the conventional method. In future work, we will investigate a method for building our empathetic DSS model in a supervised or semi-supervised setting, i.e., emotion labels of agent or user utterances are fully or partially available.

\textbf{Acknowledgements:}
This research was conducted as joint research between LINE Corporation and the Saruwatari-Koyama Laboratory of The University of Tokyo, Japan. Part of this work was supported by JSPS KAKENHI Grant Number 21K21305.

   \ninept
   \bibliographystyle{IEEEtran}
   \bibliography{template}

\end{document}